\documentclass[twocolumn]{aastex701}
\usepackage{amsmath}
\usepackage{amssymb}
\usepackage{mathtools}
\usepackage{bm}
\usepackage{tabularx}
\usepackage{booktabs}

\begin{document}

\title{Superluminal Wave Activation at Relativistic Magnetized Shocks}

\author[orcid=0000-0002-5349-7116]{Jens F. Mahlmann}
\affiliation{Department of Physics \& Astronomy, Wilder Laboratory, Dartmouth College, Hanover, NH 03755, USA}
\email[show]{jens.f.mahlmann@dartmouth.edu} 
\author[orcid=0009-0002-4241-8141]{Logan Eskildsen}
\affiliation{Department of Physics \& Astronomy, Wilder Laboratory, Dartmouth College, Hanover, NH 03755, USA}
\email[]{email}  
\author[orcid=0000-0002-3643-9205]{Arno Vanthieghem}
\affiliation{Sorbonne Université, Observatoire de Paris, Université PSL, CNRS, LUX, F-75005 Paris, France}
\email[]{email} 
\author[orcid=0009-0009-8127-8023]{Dawei Dai}
\affiliation{Department of Physics \& Astronomy, Wilder Laboratory, Dartmouth College, Hanover, NH 03755, USA}
\email[]{email}   
\author[orcid=0000-0002-1227-2754
]{Lorenzo Sironi}
\affiliation{Department of Astronomy and Columbia Astrophysics Laboratory, Columbia University, New York, NY 10027, USA}
\affiliation{Center for Computational Astrophysics, Flatiron Institute, New York, NY 10010, USA}
\email[]{email}  

\correspondingauthor{Jens F. Mahlmann}

\begin{abstract}
 Fast radio bursts (FRBs) are extremely energetic radio transients, some are generated in magnetar magnetospheres and winds. Despite a growing number of observations, their emission mechanisms remain elusive. It has recently been proposed that Alfvénic perturbations can convert into superluminal O-modes at magnetized shocks and propagate downstream as a radio signal. We validate this superluminal wave activation mechanism using pair-plasma theory and particle-in-cell simulations. Theory predicts two different downstream modes: nonpropagating Alfvénic perturbations and propagating superluminal O-modes. Superluminal wave activation occurs if the frequency of upstream perturbations in the shock frame exceeds the downstream plasma frequency. 1D particle-in-cell simulations confirm wavenumber and frequency jumps across the shock for upstream perturbations with frequencies well above the plasma frequency. Our simulations model both monochromatic upstream waves and broadband spectra with the downstream plasma frequency acting like a high-pass filter for superluminal O-modes. We discuss implications for FRB generation in relativistic magnetized winds.
\end{abstract}

\keywords{Magnetars (992); Plasma astrophysics (1261); Radio transient sources (2008); Shocks (2086); Magnetospheric radio emissions (998)}


\section{Introduction} 





Relativistic magnetized shocks generate many observable phenomena in extreme astrophysical environments, like particle acceleration to ultrahigh energies \citep[e.g.,][]{Sironi2009,Sironi2015} and 
coherent wave generation via synchrotron maser emission \citep{Iwamoto2017,Plotnikov2019,Babul2020,Sironi2021,Vanthieghem2025}. Magnetars are neutron stars with strong magnetic field strengths $B\gtrsim 10^{14}\,{\rm G}$, slow rotation periods $P\sim 1\,{\rm s}$, and coronae of magnetically active plasmas \citep{Rea2010,Kaspi2017}. Their strongly magnetized and dynamic environments are a natural place for the formation of relativistic shocks at various scales \citep[e.g.,][]{Margalit2018,Metzger2019,Beloborodov2020,Thompson2022}. Bursting, likely caused by crustal activity, can drive magnetospheric instabilities and seed magnetic fluctuations in expanding relativistic outflows.

\citet{Thompson2022} suggested that nonpropagating Alfvénic perturbations advected toward a magnetized shock convert to propagating superluminal kinetic plasma waves downstream. Such superluminal O-modes have frequencies $\omega/\bar{\omega}_{\rm p}\gtrsim 1$, where $\bar{\omega}_{\rm p}$ is the rest-frame plasma frequency \citep{Arons1986,Bransgrove2023}.
Superluminal O-modes 
could potentially escape compact object environments as radio emission, depending on the properties of the surrounding outflow.

Transient radio activity, especially fast radio bursts (FRBs), are messengers of magnetically active environments both within \citep[SGR 1935+2154;][]{Andersen2020} and outside of our Galaxy \citep{frbcollaboration2025}. These powerful millisecond-duration flares are typically narrow band (a few hundred megahertz) and their peak frequency ranges between $120\,{\rm MHz}$ and several GHz \citep{PastorMarazuela2021}. They can carry information about their host environment, for example in polarization \citep[e.g.,][]{niu2024}, and scintillation \citep{nimmo2024}. FRBs are emerging as powerful probes of turbulent plasmas throughout the Universe \citep{connor2025, Ocker2025}. However, their generation mechanisms remain elusive. This letter provides testable predictions of the \citet{Thompson2022} model for radio wave generation at relativistic shocks in compact object magnetospheres. It is organized as follows. Section~\ref{sec:activation} outlines a simplified theoretical framework with relevant scales of the wave activation process. Section~\ref{sec:simulations} shows 1D particle-in-cell (PIC) simulations (setup in Section~\ref{sec:setup}) of mode activation for monochromatic upstream fluctuations (\ref{sec:monochromatic}) and broadband seed wave spectra (\ref{sec:spectrumAW}). We discuss results and their astrophysical implications in Section~\ref{sec:discussion}.

\section{One-dimensional Model of Alfvén Wave Activation}
\label{sec:activation}

\begin{figure}
\centering
\includegraphics[width=0.475\textwidth]{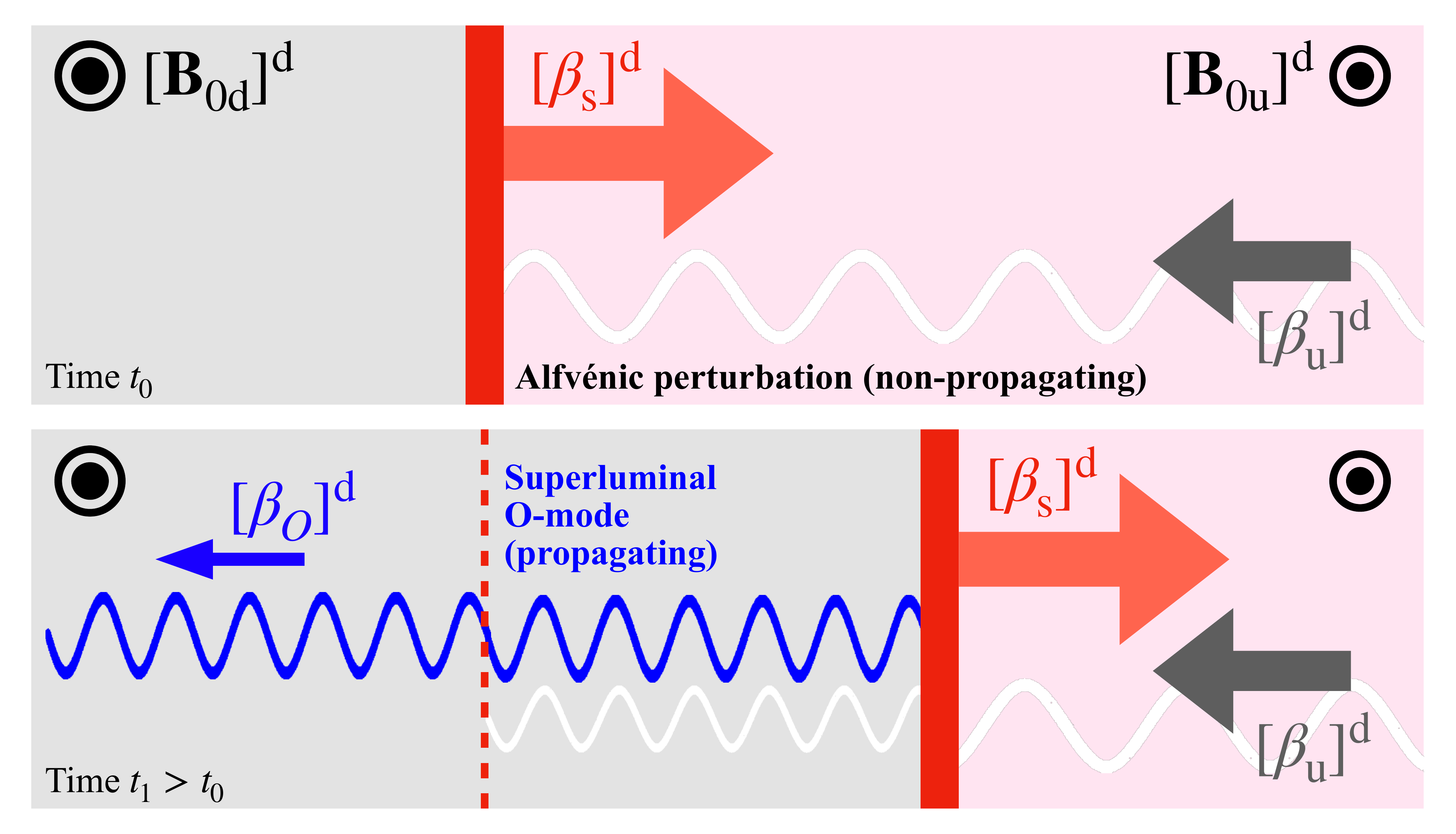}
\caption{Schematic visualization of mode conversion at relativistic magnetized shocks. A shock front propagates through a magnetized plasma. Nonpropagating Alfvénic perturbations upstream of the shock ($\omega_A=0$, waves in white) convert into a superposition of nonpropagating Alfvénic perturbations (white) and propagating superluminal O-modes (blue) in the downstream. Quantities measured in different frames are denoted as follows: 
downstream (d), upstream (u), shock (s).
}
\label{fig:figure1}
\end{figure}

We consider a 1D magnetized planar shock that propagates along $+\hat{\mathbf{x}}$ in a pair plasma with background magnetic field $[\bm{B}_{0\rm u}]^{\rm d} = \left(B_{0\rm u}/B_0\right) \hat{\mathbf{y}}$ (see Figure~\ref{fig:figure1}). Here, $eB_0=m_e\omega_{Le0}c$ normalizes the magnetic field to the electron Larmor frequency $\omega_{Le0}$ in the downstream frame. Subscripts denote to which region the quantity belongs: the downstream (d), shock front (s), or upstream (u). Superscripts indicate in which frame a quantity is measured.  This section reviews ordinary modes in cold 1D plasmas and describes a mode-conversion mechanism between subluminal (Alfvén) and superluminal modes at magnetized shocks (Figure~\ref{fig:figure1}). Nonpropagating Alfvénic perturbations frozen into the upstream flow are activated as superluminal modes propagating downstream.


\subsection{One-dimensional Ordinary Modes in Cold Plasmas}
\label{sec:dispersion}

In 1D and with  $\bm{B}_{0\rm u}$ perpendicular to $\hat{\mathbf{x}}$, only modes with wavenumber $k_\parallel = 0$ are allowed. The labels parallel ($\parallel$) and perpendicular ($\perp$) denote orientations relative to the magnetic field direction. The following derivations are for the plasma rest-frame with cold plasma dispersion relation \citep{Arons1986}:
\begin{align}
    \omega^2 (\omega^2 - c^2 k_\perp^2) (\omega^2 - c^2k_\perp^2-\omega_{\rm p}^2) = 0.
\end{align}
Here, $\omega$ is the wave frequency, and $\omega_{\rm p}=(4\pi ne^2/m_e)^{1/2}$ is the plasma frequency for a cold electron-positron pair plasma of number density $n$. The system has the following ordinary mode solutions:
\begin{align}
    \omega_A &= 0  & \text{(Alfvén)}\\
    \omega_O &= (c^2 k_\perp^2+\omega_{\rm p}^2)^{1/2} &  \text{(superluminal)}\label{eq:omode}
\end{align}
The fields $\delta\mathbf{B}$ and $\delta\mathbf{E}$ of these waves in strong magnetic fields are polarized in the $\mathbf{k}-\mathbf{B}$ plane. In the chosen 1D configuration, Alfvén waves (AWs) do not propagate. They appear as frozen-in magnetic shear fluctuations with no associated electric field, $\delta E_y=0$, and one nonvanishing component, $\delta B_z$.
The static perturbations carry a current $J_y=-(c/4\pi)\partial_x \delta B_z$ to fulfill the Ampère-Maxwell law in the plasma rest-frame. We will refer to the transverse magnetic shear fluctuations with Alfvénic polarization as \emph{Alfvénic perturbations}.

\subsection{Frequency Matching across the Shock}

Appendix~\ref{app:boosts} shows transformations of field and wave properties between relevant frames (u, s, and d). In the shock front frame, wave frequencies match across the discontinuity due to energy conservation:
\begin{align}
    [\omega_{\rm d}]^{\rm s}=[\omega_{\rm u}]^{\rm s}. \label{eq:omegatransforms}
\end{align}
\emph{Alfvénic} perturbations with $\omega_A=0$ in their respective rest-frames follow $[\gamma_{\rm s}\beta_{\rm s} k_{\rm d}]^{\rm d}=[\gamma_{\rm s}\beta_{\rm s} k_{\rm u}]^{\rm u}$ (see Appendix~\ref{app:boosts}). 
For large upstream magnetizations and relativistic flow velocities, which is typical for strong shocks in magnetized plasmas, we find
\begin{align}
    [k_{\rm d}]^{\rm d}= \frac{[\gamma_{\rm s}\beta_{\rm s}]^{\rm u}}{[\gamma_{\rm s}\beta_{\rm s}]^{\rm d}}[k_{\rm u}]^{\rm u}\approx 2[\gamma_{\rm u}]^{\rm d}[k_{\rm u}]^{\rm u}. \label{eq:udjump}
\end{align}
For \emph{superluminal} modes, frequency matching combined with Equation~(\ref{eq:omode}) implies a downstream wavenumber
\begin{align}
    c^2[k^2_{\rm d}]^{\rm s} = 
    [\omega_{\rm u}^2]^{\rm s} - \omega_{\rm pd}^2= c^2[\gamma_{\rm s}^2\beta_{\rm s}^2 k_{\rm u}^2]^{\rm u} - \omega_{\rm pd}^2\label{eq:omodematching}
\end{align}
Here, $\omega_{\rm pd}$ is the plasma frequency calculated with the downstream density, and we evaluated the dispersion relation in the shock-front frame (Equation~\ref{eq:omodeshock}). When $[\omega_{\rm u}]^{\rm s}<\omega_{\rm pd}$, no propagating modes exist. In the high-frequency limit $[\omega_{\rm u}]^{\rm s}\gg\omega_{\rm pd}$, the downstream wavenumber can be approximated as (see Appendix~\ref{app:boosts}):
\begin{align}
    [k_{\rm d}]^{\rm d} =  2[\gamma_{\rm d}]^{\rm s}[\gamma_{\rm s} \beta_{\rm s} k_{\rm u}]^{\rm u}\approx [\gamma_{\rm u}]^{\rm d}[k_{\rm u}]^{\rm u}\label{eq:omodejump}.
\end{align}
The last approximation again requires high flow velocities in the shock front frame. Equation~(\ref{eq:omodematching}) constrains the minimum wavenumber of seed perturbations to allow for downstream-propagating modes: 
\begin{align}
 \frac{c[k_{\rm u}]^{\rm u}}{\omega_{\rm pu}}[\gamma_{\rm s}\beta_{\rm s}]^{\rm u}\ge \frac{\omega_{\rm pd}}{\omega_{\rm pu}}\approx\left(2[\gamma_{\rm u}]^{\rm d}\right)^{1/2}
 \label{eq:mink}
\end{align}
Here, $\omega_{\rm pu}$ is the plasma frequency calculated with the upstream density. We examine the scales for propagating and nonpropagating waves across the shock normalized to the respective plasma skin depth and frequency. For the upstream wavenumber normalized by a plasma skin depth $[d_{\rm u}]^{\rm d}=[d_{\rm u}]^{\rm u}/[\gamma_{\rm u}]^{\rm d}=c/(\omega_{\rm pu}[\gamma_{\rm u}]^{\rm d})$, we find
\begin{align}
    [k_{\rm u} d_{\rm u}]^{\rm d} = [k_{\rm u} d_{\rm u}]^{\rm u}. \label{eq:kdmatching}
\end{align}
Combining Equation (\ref{eq:omegatransforms}) with Equations~(\ref{eq:udjump}) and~(\ref{eq:omodematching}) then implies the following jumps in normalized wavenumber between upstream perturbations and downstream modes across the shock:
\begin{align}
[k_{\rm d}d_{\rm d}]^{\rm d} = \sqrt{2[\gamma_{\rm u}]^{\rm d}}\, [k_{\rm u}d_{\rm u}]^{\rm u} & \quad \text{(Alfvén)}\label{eq:AWUpMatch}\\
[k_{\rm d}d_{\rm d}]^{\rm d} = \sqrt{\frac{[\gamma_{\rm u}]^{\rm d}}{2}} \, [k_{\rm u}d_{\rm u}]^{\rm u} & \quad \text{(superluminal)} \label{eq:DownUpMatch}
\end{align}
Here, we used $[k_{\rm u}]^{\rm d} = [k_{\rm d}]^{\rm d}/[\beta_{\rm s}]^{\rm u}$ (see Appendix~\ref{app:frames}) in the limit of $c[k_{\rm d}]^{\rm d}\gg \omega_{\rm pd}$.

\subsection{Amplitude Matching across the Shock}

Nonpropagating Alfvénic perturbations on a constant background $B_{0 \rm u}$ in the upstream rest-frame have electromagnetic (EM) fields
\begin{align}
    [\mathbf{E}_{\rm u}]^{\rm u} = [(0, 0, 0)]^{\rm u}\qquad [\mathbf{B}_{\rm u}]^{\rm u} = [(0, B_{0 \rm u}, \delta B_{z\rm u})]^{\rm u}\label{eq:upstreamfieldsU}.
\end{align}
MHD jump conditions (Appendix~\ref{app:jumps}) imply that electric fields are continuous in the frame of the shock. For relativistic shocks, the magnetic field experiences the compression $[B_{y\rm d}]^{\rm d}/[B_{y\rm u}]^{\rm d}\approx 2$. Upstream waves at MHD scale 
will experience a similar field compression, $[\delta B_{z\rm d}]^{\rm d}/[\delta B_{z\rm u}]^{\rm d}\approx 2$. High-frequency waves with $[\omega_{\rm u}]^{\rm d}\gtrsim\omega_{\rm pu}[\gamma_{\rm u}]^{\rm d}$ essentially propagate as EM waves with $[\delta B_{z\rm d}]^{\rm d}/[\delta B_{z\rm u}]^{\rm d}\approx 1$ and continuous Poynting flux across the shock.
Then, EM fields are continuous in all frames. In the downstream frame
\begin{align}
    [E_{y\rm d}]^{\rm d}=[E_{y\rm u}]^{\rm d}=-[\gamma_{\rm d} \beta_{\rm d} \delta B_{z\rm u}]^{\rm u}.
\end{align}
In the high-frequency limit, the only mode carrying this electric field in the downstream frame is the superluminal O-mode with dispersion as in Equation~(\ref{eq:omode}). Using the group velocity $\beta_{O}=ck_\perp/\omega_O$ and the Maxwell-Faraday equation, $\omega\,\delta B_z=ck_\perp \delta E_y$, we find the magnetic field of superluminal modes propagating along $-\hat{\mathbf{x}}$:
\begin{align}
    [\delta B_{O\rm d}]^{\rm d}=[E_{y\rm d}\beta_{O\rm d}]^{\rm d}=-[\gamma_{\rm d} \beta_{\rm d} \delta B_{z\rm u}]^{\rm u}[\beta_{O\rm d}]^{\rm d}.
\end{align}
Assuming that the total downstream magnetic field is shared between modes, we estimate the amplitude of the Alfvénic perturbation for high-frequency seed waves:
\begin{align}
    [\delta B_{A\rm d}]^{\rm d}/[\delta B_{z\rm u}]^{\rm d}\approx 1+[\beta_{\rm d}]^{\rm u}[\beta_{O\rm d}]^{\rm d}.
     \label{eq:AWampD}
\end{align}
For $c[k_{\rm d}]^{\rm d}\gg \omega_{\rm pd}$, the wave speed $\beta_O\approx -1$ (Equation~\ref{eq:omode}), such that $[\delta B_{A\rm d}]^{\rm d}\rightarrow 0$. At high frequencies, only superluminal O-modes are generated downstream.



\section{Simulations}
\label{sec:simulations}

We conduct 1D PIC simulations of mode conversion at a magnetized relativistic perpendicular shock in pair plasma with the \textsc{Tristan-MP.v2} code \citep{tristanv2}. 

\subsection{Setup}
\label{sec:setup}

Our simulation setup follows previous simulations of relativistic collisionless shocks \citep[e.g.,][]{Spitkovsky2008,Sironi2009,Parsons2024}. We initialize an upstream flow of magnetized pair plasma with Lorentz factor $[\gamma_{\rm u}]^{d}$ along the $-\mathbf{\hat{x}}$ direction and a thermal spread $T/m_e c^2=10^{-3}$ in units with $k_B=1$. The magnetic field is initially uniform with $[\bm{B}_{0\rm u}]^{\rm d} = \left(B_{0\rm u}/B_0\right) \hat{\mathbf{y}}$. A conducting wall located at $x=0$ reflects particles and acts as a conducting boundary for fields. The counterstreaming flows form a shock that propagates along the $+\mathbf{\hat{x}}$ direction. The simulation frame is the downstream rest-frame. We simulate a domain with $N=96\times 10^4$ cells, 
and a skin depth $[d_{\rm u}]^{\rm d}=240$ cells. The total fixed-length domain extends for $N/[d_{\rm u}]^{\rm d}=4\times 10^3$ upstream plasma skin depths. We choose 
an upstream magnetization $[\sigma_{\rm u}]^{u}= [B_{0\rm u}^2/\gamma_{\rm u}]^{\rm d}$. The initial flow fills a region extending up to $x/[d_{\rm u}]^{\rm d}\lesssim 40$, and fields and particles are continuously replenished by a moving injector located upstream of the shock. Once the shock has fully developed and propagates in the upstream plasma, the injector additionally generates wave fields in the simulation frame, as outlined below. Currents to support this perturbation are supplied by a particle velocity component
along $\mathbf{\hat{y}}$ (see Section~\ref{sec:dispersion}). We track wave and shock dynamics for various properties of the upstream flow and upstream seed modes.

\begin{figure*}
\includegraphics[width=0.495\textwidth]{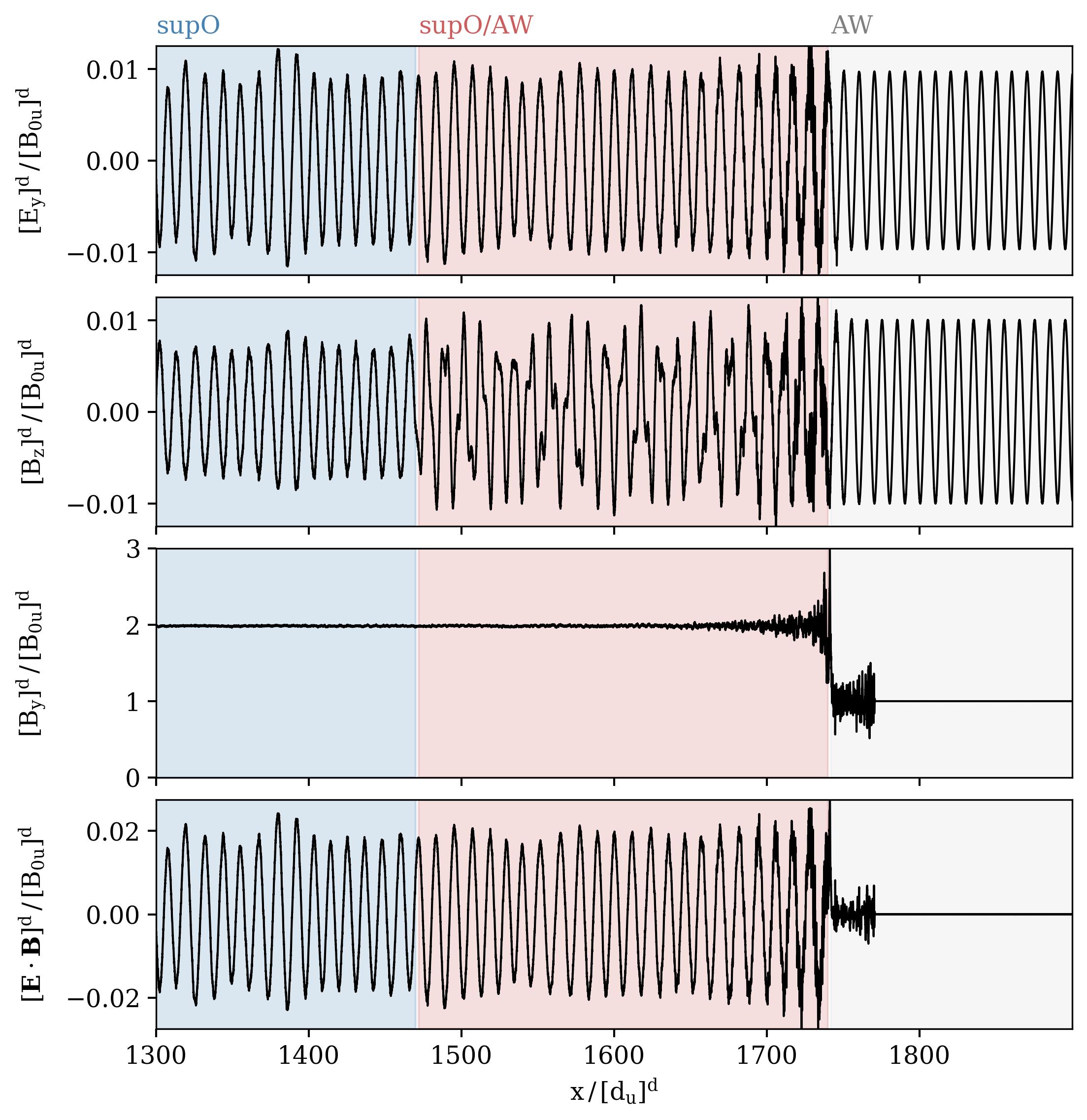}
\includegraphics[width=0.495\textwidth]{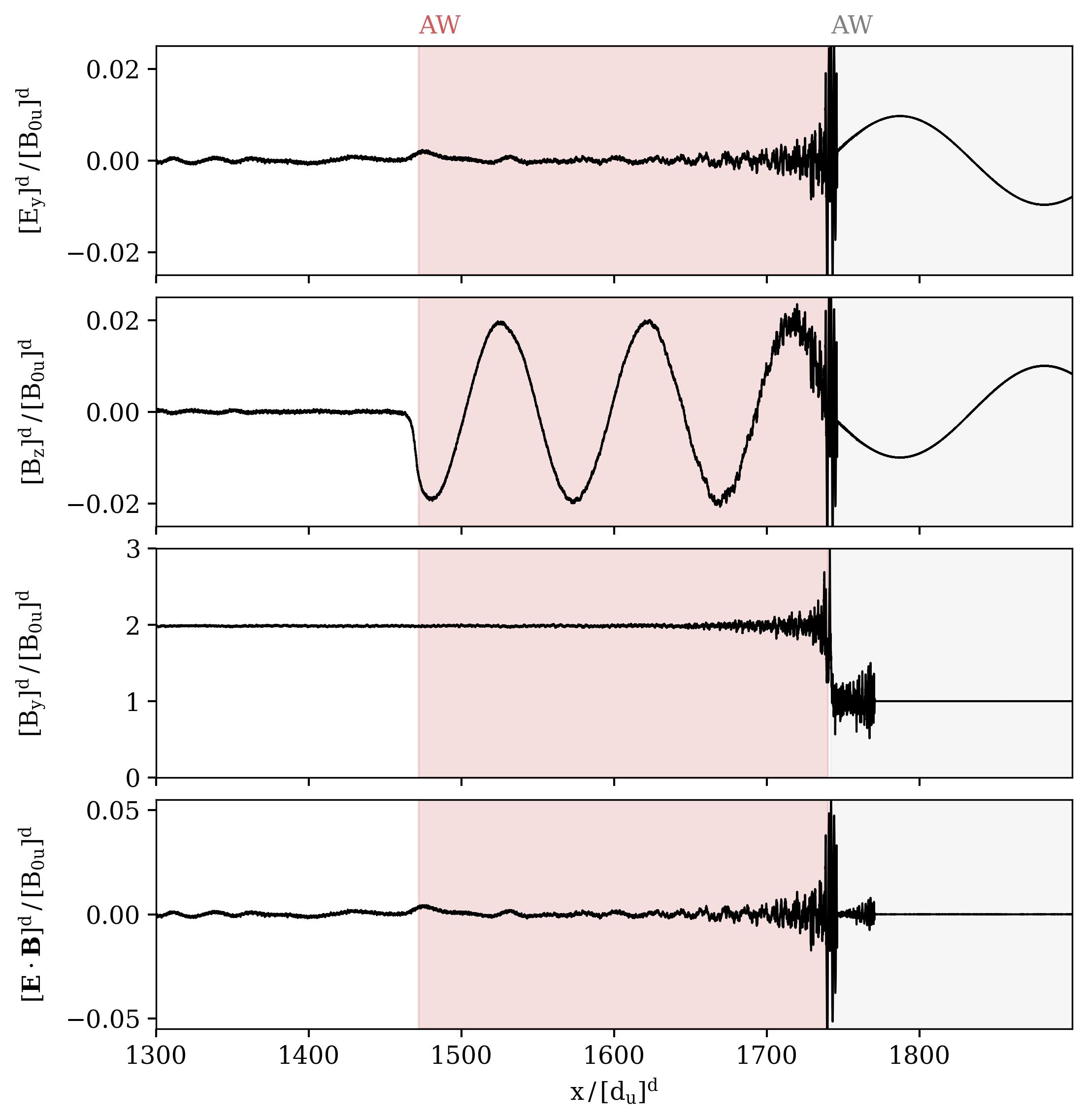}\\
\includegraphics[width=0.495\textwidth]{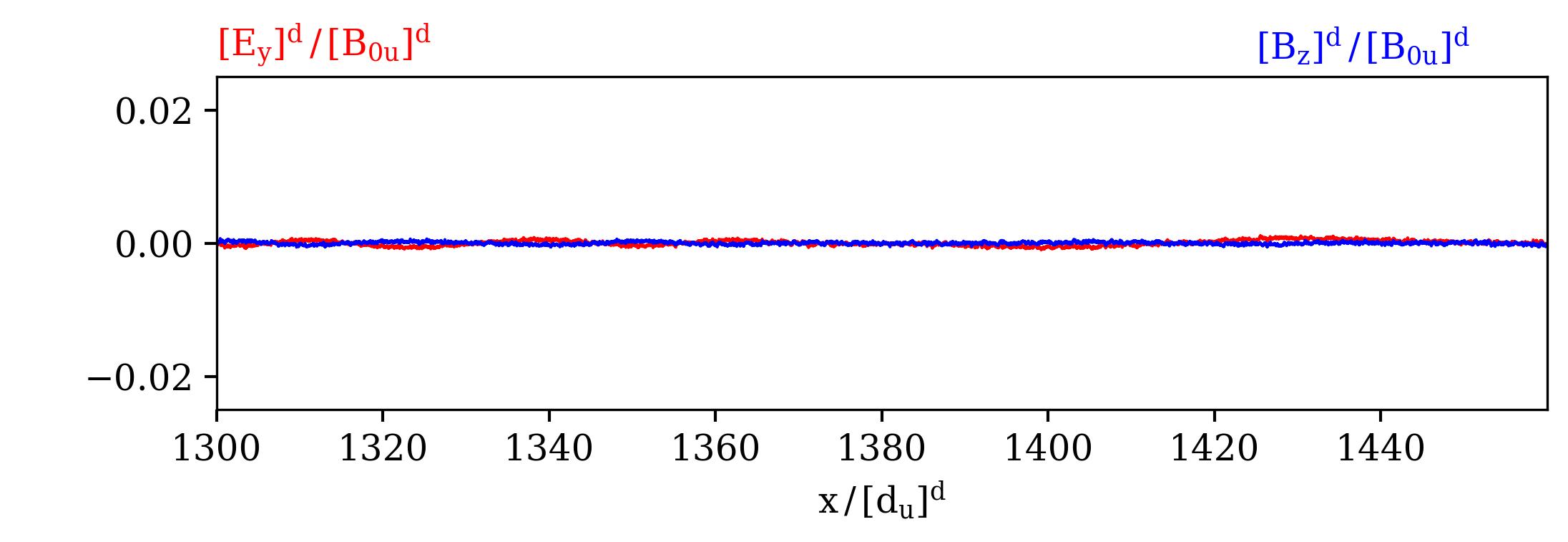}
\includegraphics[width=0.495\textwidth]{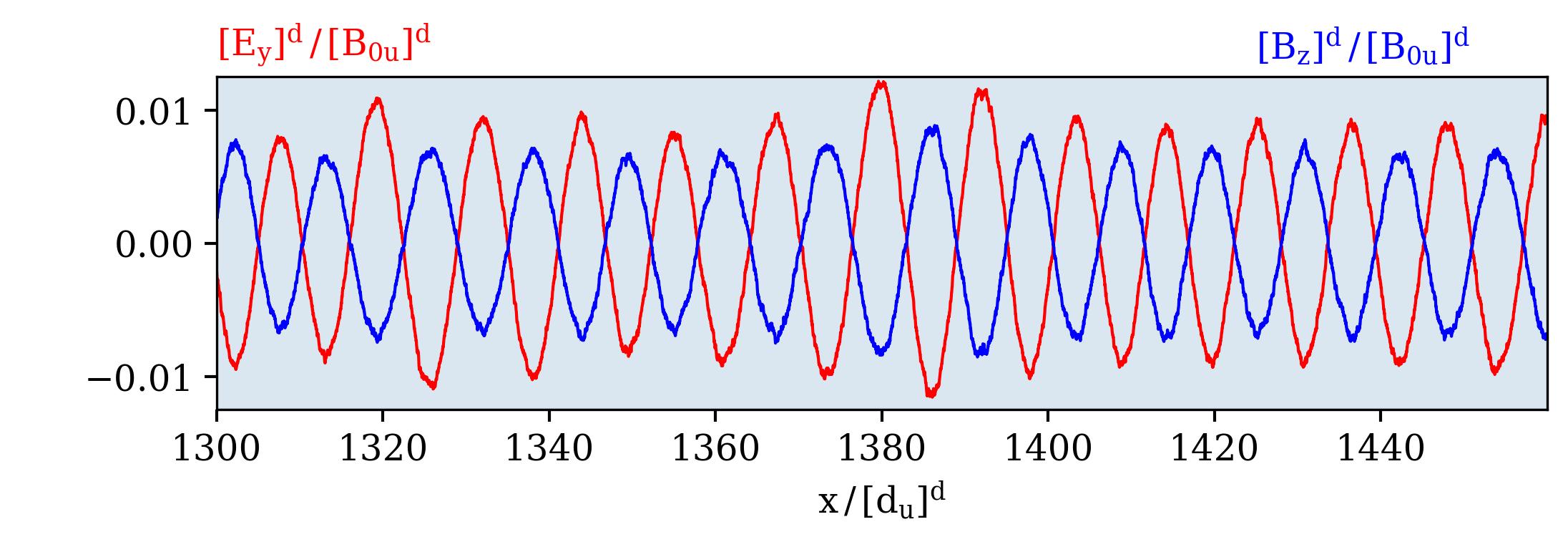}
\vspace{-6pt}
\caption{Shock and mode-conversion dynamics of monochromatic seed waves. Left: upstream scales $B_0=10$, $[\gamma_{\rm u}]^{\rm d}=4$, $[k_{\rm u}d_{\rm u}]^{\rm u}=0.63$. A superluminal O-mode (supO) propagates downstream, a small-amplitude nonpropagating Alfvénic perturbation (AW) remains in the wake of the shock. The region of pure supO is shaded in blue, mixed supO/AW in red. The bottom panel shows EM fields associated with the propagating superluminal O-mode. Right: upstream waves with $[k_{\rm u}d_{\rm u}]^{\rm u}=0.03$. No modes propagate downstream, only AW perturbations remain (red shaded region of top panels). We provide online animations of the wave activation dynamics in the supporting material \citep[see][]{SupplementaryMediaA,SupplementaryMediaB}.}
\label{fig:MSHOCK_G4_s25_HR_K0026}
\end{figure*}

    \begin{figure*}
    \centering
\includegraphics[width=0.49\textwidth, height=8.1cm, keepaspectratio]{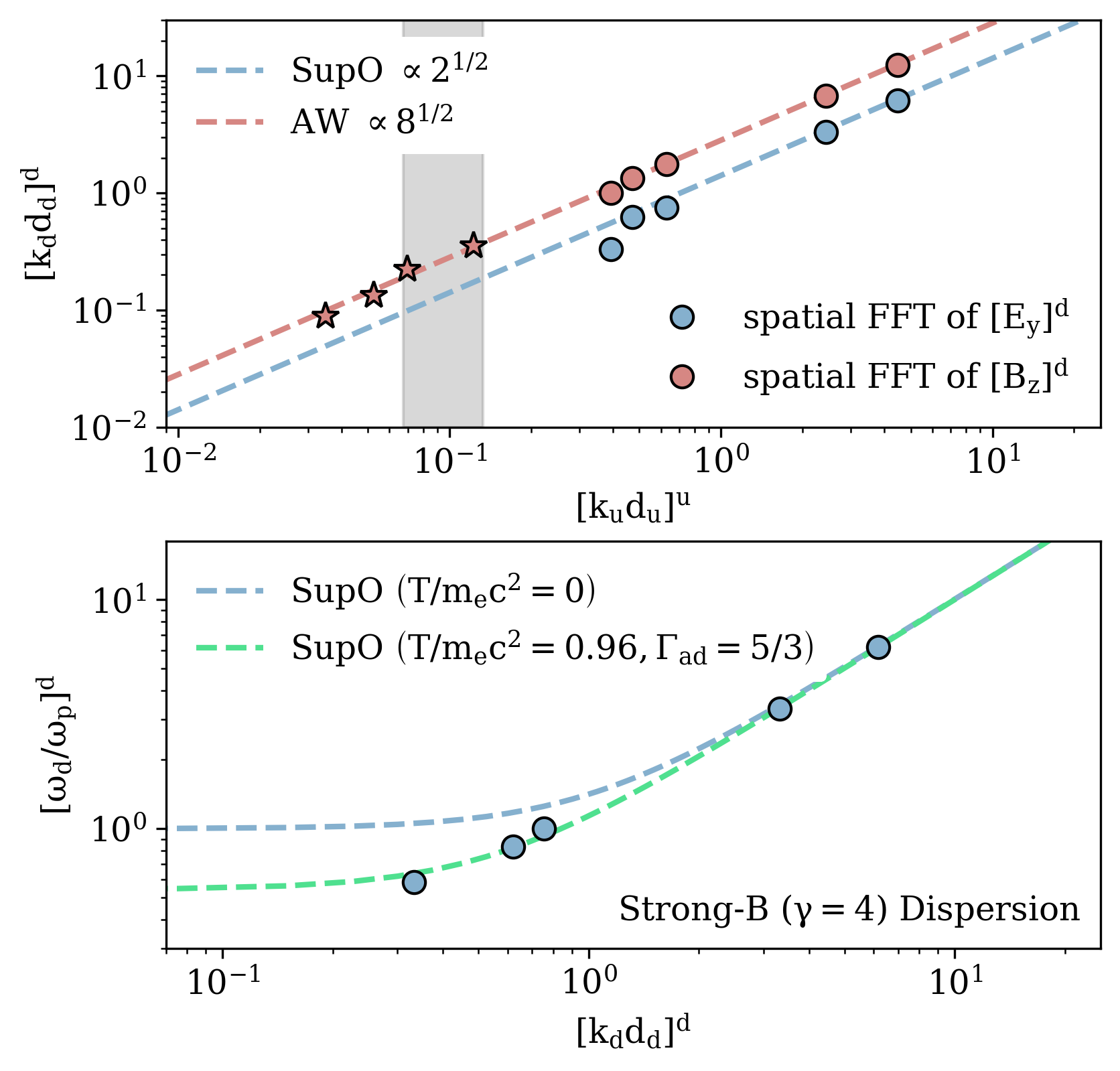}
\includegraphics[width=0.49\textwidth, height=8.1cm, keepaspectratio]{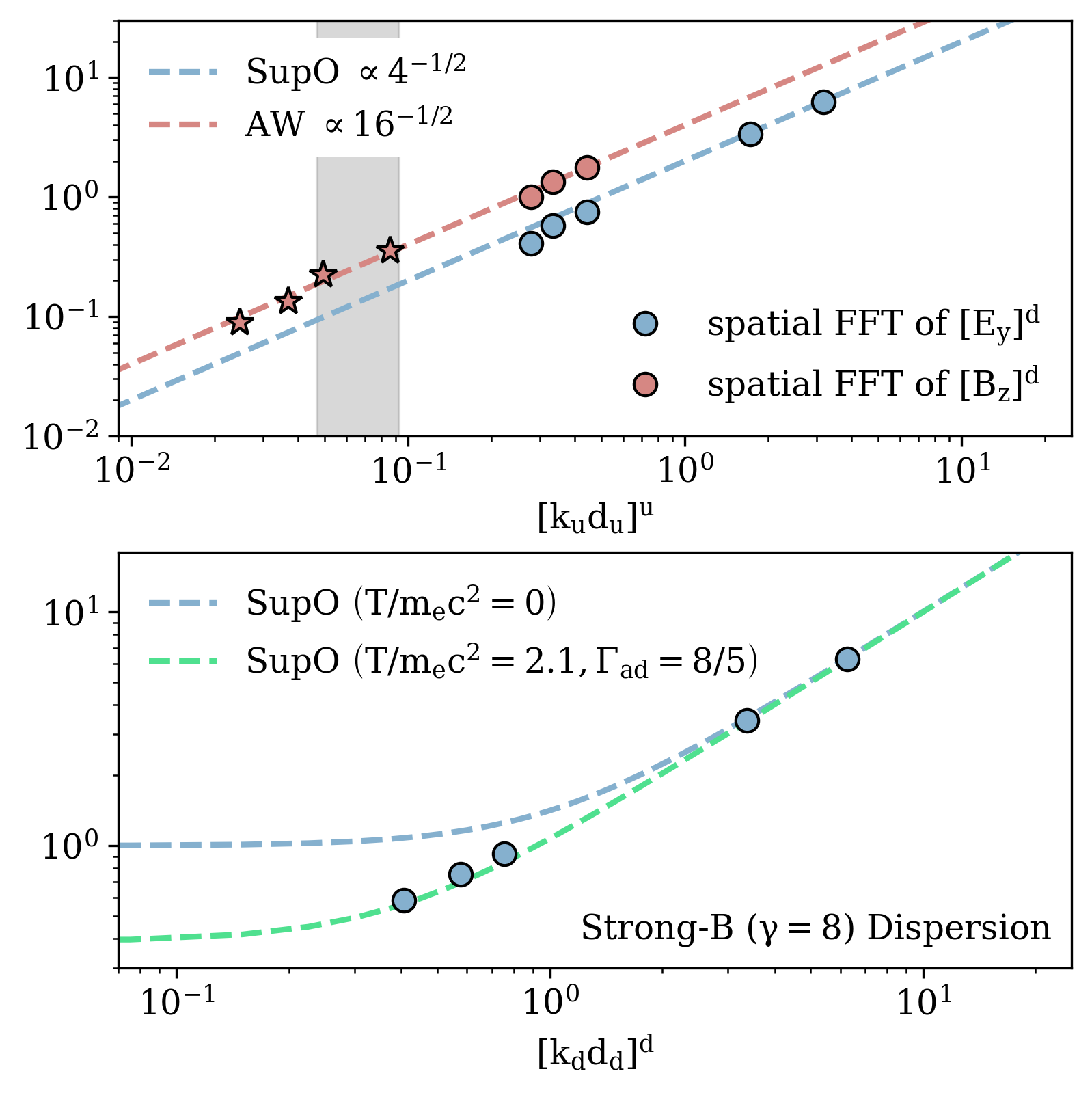}
    \caption{Wavenumber scaling (top) and frequency matching (bottom) across shocks with $[\gamma_{\rm u}]^{\rm d} = 4$ (left) and $[\gamma_{\rm u}]^{\rm d} = 8$ (right). Top panels show downstream wavenumbers as a function of upstream seed wavenumbers in the superposition region for AWs (red circles) and superluminal O-modes (blue circles). The scalings agree closely with Equations~(\ref{eq:AWUpMatch}) and (\ref{eq:DownUpMatch}). Star markers denote cases for which no propagating downstream modes can be measured. Gray shading indicates the range between the plasma-frequency cutoff estimates derived in the cold plasma and warm plasma limits (using Equation~\ref{eq:omodematching}). Bottom panels show the dispersion relation of propagating downstream modes. Downstream wavenumbers and corresponding wave frequencies trace the respective (hot) dispersion relation for the SupO-mode with temperatures $[T_{\rm d}]^{\rm d} = 0.96\, m_ec^2$ for $[\gamma_{\rm u}]^{\rm d} = 4$, and $[T_{\rm d}]^{\rm d}=2.1\, m_e c^2$ for $[\gamma_{\rm u}]^{\rm d} = 8$ (see also Equation~\ref{eq:HPD}).}
    \label{fig:figure4}
    \end{figure*}

\subsection{Dependence of Upstream Scales}
\label{sec:monochromatic}

We first initialize monochromatic Alfvénic perturbations upstream of the shock, corresponding to
\begin{align}
[\mathbf{B}_{\rm u}]^{\rm d} &= \left(0, [B_{0\rm u}]^{\rm d}, a_0 [B_{0\rm u}]^{\rm d}\times\cos\left([k_{\rm u}]^{\rm d}x\right)\right)\label{eq:upstreamfieldsU}\\
[\mathbf{E}_{\rm d}]^{\rm u}&=-[\boldsymbol{\beta}_{\rm u}]^{\rm d}\times[\mathbf{B}_{\rm u}]^{\rm d}\label{eq:waveEfield},
\end{align}
where $a_0$ is a dimensionless amplitude parameter and we use $a_0=10^{-2}$ in this work. We probe the limits of AW activation for $[\gamma_{\rm u}]^{\rm d}=4$ and an upstream magnetic field of $[\mathbf{B}_{\rm u}]^{\rm d}/B_0=10$. 
Figure~\ref{fig:MSHOCK_G4_s25_HR_K0026} (left panel) displays a setup that initializes Alfvénic perturbations at kinetic scales in the upstream plasma, using $[k_{\rm u}d_{\rm u}]^{\rm u}=0.63$. This implies $[\omega_{\rm u}]^{\rm s}/(\bar{\omega}_{\rm pd}/[\gamma_{\rm u}]^{\rm d})\approx 18.37\gg 1$, where we approximate the hot downstream plasma frequency for adiabatic index $\Gamma_{\rm ad}$ and temperature $T$ as 
\begin{align}
    \bar{\omega}_{\rm pd}={\omega}_{\rm pd}\times\left(1+\frac{\Gamma_{\rm ad}}{\Gamma_{\rm ad} - 1}\frac{[T_{\rm d}]^{\rm d}}{m_e c^2}\right)^{-1/2}. \label{eq:HPD}
\end{align}
We measure the downstream temperature $[T_{\rm d}]^{\rm d}/m_ec^2 = [\left\langle \gamma_{\rm d} \beta_{\rm d}^2\right\rangle/3]^{\rm d} = 0.96$, where $\left\langle\dots\right\rangle$ averages individual particle velocities. We use the adiabatic index $\Gamma_{\rm ad} = ( 2 + 3 \,T/m_ec^2 )/( 1 + 2\,T/m_ec^2)$ appropriate for 2D Maxwellian distributions, and approximate $T/m_ec^2$ by the nearest integer. Incoming waves couple to superluminal modes downstream of the shock. For these scales, Equation~(\ref{eq:AWampD}) suggests small amplitudes of residual AWs in the downstream with $[\delta B_{A\rm d}]^{\rm d}/[\delta B_{z\rm u}]^{\rm d}\approx 0.03$ for superluminal O-modes propagating downstream with a group velocity close to the speed of light. These features are consistently reproduced by the simulation. 
Two regions form downstream: the far downstream region which is reached only by the propagating superluminal O-modes (shaded blue in Figure~\ref{fig:MSHOCK_G4_s25_HR_K0026}, left panel), and the immediate wake of the shock showing a superposition of superluminal O-modes and nonpropagating small-amplitude Alfvénic perturbations (shaded red in Figure~\ref{fig:MSHOCK_G4_s25_HR_K0026}, left panel). In the downstream region, $\mathbf{E}\cdot\mathbf{B}$ is generated at the scale of the propagating O-mode, as would be expected from plasma dispersion \citep[][]{Arons1986,Bransgrove2023}.

Figure~\ref{fig:MSHOCK_G4_s25_HR_K0026} (right panel) analyses the same shock and flow properties, but for significantly longer upstream wavelengths $[k_{\rm u}d_{\rm u}]^{\rm u}=0.03$. Then, $[\omega_{\rm u}]^{\rm s}/(\bar{\omega}_{\rm pd}/[\gamma_{\rm u}]^{\rm d})\approx 0.97 \lesssim 1$, and Equation~(\ref{eq:omodematching}) implies no coupling to superluminal O-modes downstream of the shock. The simulation data confirm this prediction. Only nonpropagating AWs remain in the immediate downstream, and residual O-mode perturbations with negligible amplitudes and varying wavelengths enter the downstream region. Also, the MHD-scale upstream mode amplitude increases by a factor of approximately two as expected from MHD jump conditions (Appendix~\ref{app:jumps}). 

\subsubsection{Dispersion and Wave Scales}

Figure \ref{fig:figure4} analyzes simulations for varying upstream wavenumber $[k_{\rm u}]^{\rm u}$ and upstream velocity $[\gamma_{\rm u}]^{\rm d}$, with the other parameters as in Section~\ref{sec:monochromatic}. Top panels show Fourier analysis in space for $[\gamma_{\rm u}]^{\rm d} = 4$ and $[\gamma_{\rm u}]^{\rm d} = 8$. Bottom panels probe dispersion through temporal FFT at a fixed coordinate in the downstream region. From each simulation, we combine results from four independent  FFTs: 1) $E_y$ spatially in the supO region (blue circles) 2) $B_z$ spatially in the supO/AW region (red circles), 3) $E_y$ spatially in the upstream region (red/blue circles) and 4) $E_y$ temporally in the supO region to capture the frequency of the downstream propagating modes. We then compare the measured wavenumber to Equations~(\ref{eq:AWUpMatch}) and~(\ref{eq:DownUpMatch}). The predictions capture the downstream wavenumbers well for both nonpropagating AWs and propagating modes. 

As discussed in Equation~(\ref{eq:HPD}), estimates of the plasma frequency must consider thermal effects for meaningful normalization of the downstream frequency. We measure downstream temperatures of $[T_{\rm d}]^{\rm d} = 0.96\,m_ec^2$ for $[\gamma_{\rm u}]^{\rm d} = 4$ and $[T_{\rm d}]^{\rm d}=2.10\,m_ec^2$ for $[\gamma_{\rm u}]^{\rm d} = 8$. This allows us to compare the cold plasma dispersion $T=0$ (dashed blue lines) to the realistic plasma dispersions (dashed green lines) in the bottom panels (b). The measured propagating wave frequencies closely fit the expectation from Equation~(\ref{eq:omodematching}) with corrections for thermal effects. Estimating the plasma‑frequency cutoff (Equation~\ref{eq:omodematching}) using the hot‑plasma frequency shifts the activation limit toward higher wavenumbers. In Figure~\ref{fig:figure4}, we shade in gray the region between the cutoff frequencies derived for hot and cold plasmas.

The results from the parameter scan in Figure~\ref{fig:figure4} confirm the frequency and wavenumber matching relationships derived in Section \ref{sec:activation}. We will explore other parameter space dimensions like magnetization and wave-plasma resonances in the future (see Section \ref{sec:discussion}).

\begin{figure*}
\includegraphics[width=1\textwidth]{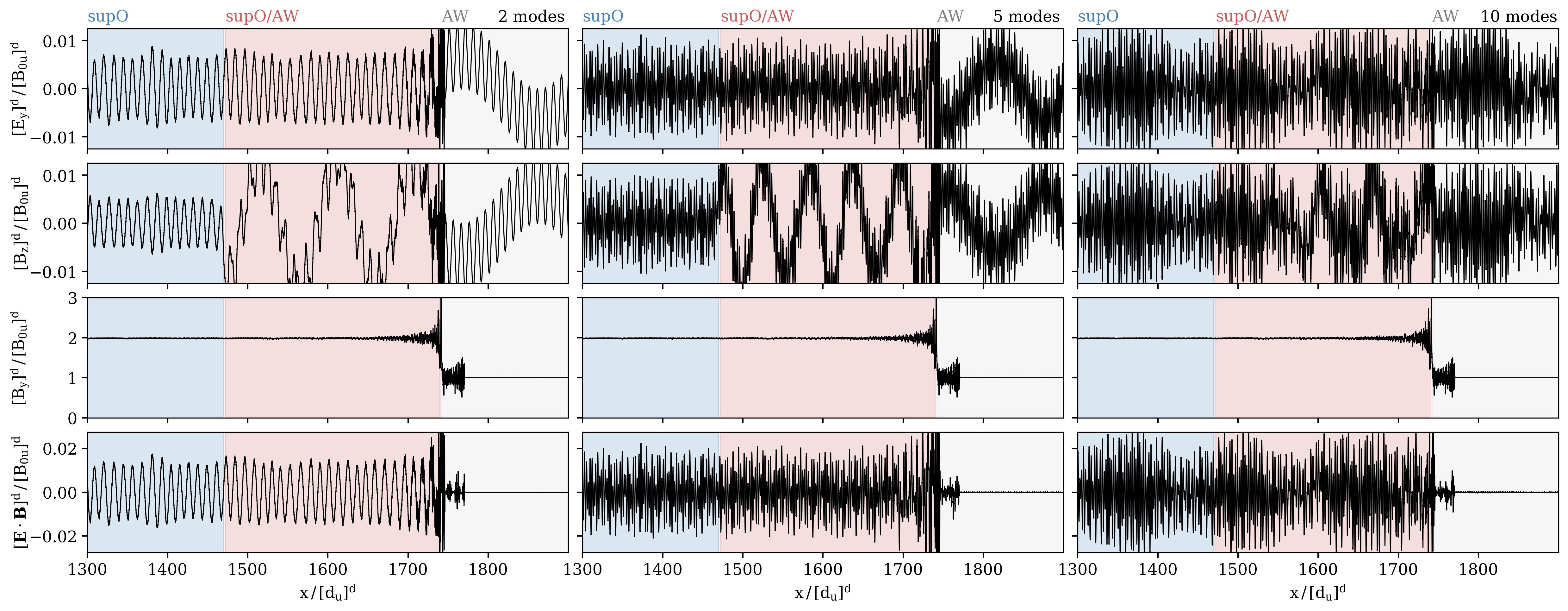}\\
\includegraphics[width=1\textwidth]{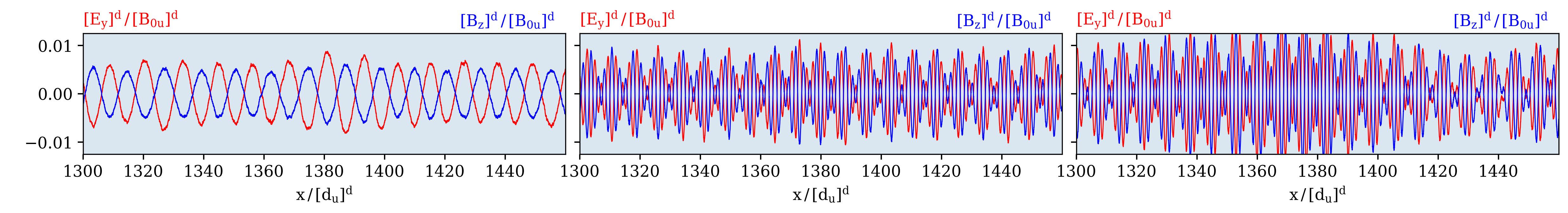}
\vspace{-16pt}
\caption{Shock and mode conversion dynamics for a (discrete) spectrum of seed waves. Flow properties are as in Figure~\ref{fig:MSHOCK_G4_s25_HR_K0026}. We initialize a spectrum of equal amplitude monochromatic waves sampled from $[k_{\rm u}d_{\rm u}]^{u}\in\left[0.03,6.28\right]$. We vary the number of modes; the left panel uses the minimum and maximum seed wavelength from Figure~\ref{fig:MSHOCK_G4_s25_HR_K0026} ($N=2)$, the middle/right have randomly sampled seed wavenumbers ($N=5$ and $N=10$, in log space). Low-frequency seed waves produce nonpropagating downstream AWs (red shaded region). Seed waves with scales above the critical frequency (Equation~\ref{eq:omodematching}) induce propagating superluminal O-modes (blue shaded region). The bottom panel zooms in on the wave fields $[E_y]^{\rm d}$ (red color) and $[B_z]^{\rm d}$ (blue shaded regions).}
\label{fig:MSHOCK_G4_s25_HR_sPEC10}
\end{figure*}

\subsection{Activation of a Spectrum of Seed Perturbations}
\label{sec:spectrumAW}

For this section, we initialize a distribution of seed perturbations resembling turbulent fluctuations upstream of the shock. For a given range of wavenumbers between $[k_{\rm u,min}]^{\rm d}$ and $[k_{\rm u,max}]^{\rm d}$ we generate a spectrum of $N$ monochromatic waves with random wavenumbers $[k_{i,\rm u}]^{\rm d}$ (lognormal distribution), amplitudes $A_i$ (Gaussian noise distribution), and phases $\phi_i$ (uniform distribution). We normalize individual wave amplitudes with a power-law spectral envelope such that their energy spectrum is approximately $e(k)\propto A(k)^2\propto k^{-s}$ for a spectral index $s$ and the total energy across all modes is $E_{\rm tot}$. The wave magnetic field then becomes
\begin{align}
    [\delta B_{z,\rm u}]^{\rm d}=\sum_{i=1}^N A_i\times\cos\left([k_{i,\rm u}]^{\rm d}x+\phi_i\right).
\end{align}
The electric field follows from Equation~(\ref{eq:waveEfield}), currents are initialized as outlined in Section~\ref{sec:monochromatic}. 

Figure~\ref{fig:MSHOCK_G4_s25_HR_sPEC10} shows shock and mode-conversion dynamics for seed waves sampled of equal amplitude ($s=0$) from $[k_{\rm u}d_{\rm u}]^{\rm u}\in\left[0.03,6.28\right]$. For illustration purposes, we choose $N=2$, $N=5$ and $N=10$ superimposed monochromatic modes. The case of $N=2$ combines the two upstream scales displayed in Figure~\ref{fig:MSHOCK_G4_s25_HR_K0026}. Large-scale fluctuations with wavelength above the critical scale defined by Equation~(\ref{eq:mink}) become nonpropagating AWs downstream of the shock. High-frequency components of the seed spectrum are activated to become propagating O-modes, the shock acts like a high-pass filter. The different components can be identified in the blue (supO) and red (supO/AW) shaded regions of Figure~\ref{fig:MSHOCK_G4_s25_HR_sPEC10}. Appendix~\ref{app:spectra} discusses the spectra of seed waves with $s\approx 1$, with the overall energy distribution of downstream modes following the slope of the seed wave spectrum.


\section{Discussion and Conclusions}
\label{sec:discussion}
Ultramagnetized pair-dominated plasmas in compact object magnetospheres allow three different waves: subluminal O-modes (AWs), superluminal O-modes, and X-modes \citep{Arons1986}. Each can have different generation mechanisms and propagation behavior. AWs can be generated by plasma instabilities \citep{Usov1987,zeng2025}, and magnetic field line motion on the magnetar crust \citep{Carrasco2019,Yuan2020,Yuan2021,Yuan2022,Chen2025,Mahlmann2023,Mahlmann2024}. AWs propagate along magnetic field lines and experience Landau damping, confining their dynamics to the magnetically active inner magnetosphere. Linear X-modes generally do not couple to plasma and can freely propagate in the magnetosphere as EM waves. 
Long-wavelength X-mode waves can steepen on decaying magnetic fields and generate shocks \citep{chen2022,Beloborodov2023,Vanthieghem2025}. Superluminal O-modes do not experience Landau damping and can escape the magnestosphere similar to X-modes. However, plasma properties affect their propagation. 

Although superluminal O-modes propagate like an EM wave at high frequencies, they have different polarization than X-modes, with
different electric field orientation toward the $\boldsymbol{k}-\boldsymbol{B}$ plane. Polarimetric observations indicating circular polarization in some FRBs \citep[e.g.,][]{Uttarkar2024} could, among other possibilities, suggest that both X- and O-modes contribute to the escaping radio signal. This letter discusses a mechanism of O-mode activation from Alfvénic perturbations at a relativistic shock. In this scenario, AWs escape the inner magnetosphere by propagating into a magnetized outflow. In the relativistic plasma of the spherically expanding wind, their $k_\parallel$ vanishes, essentially `freezing' the AWs into the flow \citep[see discussion by][]{Thompson2022}. At least three types of waves can be generated at the shock: propagating high-frequency superluminal O-modes, low-frequency Alfvénic perturbations, and X-modes via synchrotron maser emission. A first-principles plasma treatment including all interacting modes is essential for a realistic description of radio wave generation at relativistic shocks.

\subsection{Process Efficiency and Frequency Range}

Throughout this discussion, the wavenumber cutoff is defined using the cold plasma frequency. We intentionally adopt the cold plasma approximation to obtain conservative lower limits on the cutoff, since the electron temperature in many of the targeted astrophysical environments is poorly constrained. Hot-plasma corrections generally shift the cutoff to smaller scales (higher wavenumbers). In the simulations presented in Section~\ref{sec:simulations}, where the temperature is explicitly known, we verified this expectation by benchmarking against the corresponding hot-plasma frequency.

The shock strength parameter $\mathcal{S}$ is commonly quantified by the ratio of Lorentz factors of upstream and downstream flows measured in the shock frame (Appendix~\ref{app:ShockStructure}). Large values corresponds to more compressive and energetic shocks. The wavenumber limit in Equation~(\ref{eq:mink}) requires a minimum upstream scale
\begin{align}
    [k_{\rm u}d_{\rm u}]^{\rm u}\gtrsim \left(\mathcal{S}[\sigma_{\rm u}]^{\rm u}\right)^{-1/2}
    \label{eq:downstreamcritical}
\end{align}
to seed downstream-propagating modes. Strong magnetized shocks allow for a larger range of upstream fluctuations to be activated. We approximate the resulting downstream scale of superluminal waves via Equation~(\ref{eq:omodejump}):
\begin{align}
    [k_{\rm d}d_{\rm d}]^{\rm d}\gtrsim 
    \left(\frac{1+\mathcal{S}^2}{4\mathcal{S}^2[\sigma_{\rm u}]^{\rm u}}\right)^{1/2}\approx \frac{1}{2}\left([\sigma_{\rm u}]^{\rm u}\right)^{-1/2}.
    \label{eq:finalcutoff}
\end{align}
In this mechanism, interactions of nonpropagating AWs with relativistic magnetized shocks drive the activation of superluminal O-modes. Linear coupling generates propagating waves with amplitudes that match the incoming seed waves at scales well above the cutoff frequency in Equation~(\ref{eq:mink}). In the analyzed parameter range, we do not find additional coherent generation, resonant enhancement, or suppression of superluminal O-modes. 
Defining the efficiency of radio wave generation by this process requires the context of global models. It depends on the fraction of available energy that can be transported to the shock location at fluctuation scales that will couple to propagating O-modes in the downstream. 
The bandwidth of generated superluminal modes is limited at low frequencies by the plasma-frequency cutoff in Equation~(\ref{eq:finalcutoff}). At high frequencies, the signal band extends to the scale of the highest-frequency seed fluctuations. 

\subsection{Signals in the Observer Frame}

    \begin{figure}
    \centering
    \includegraphics[width=0.475\textwidth]{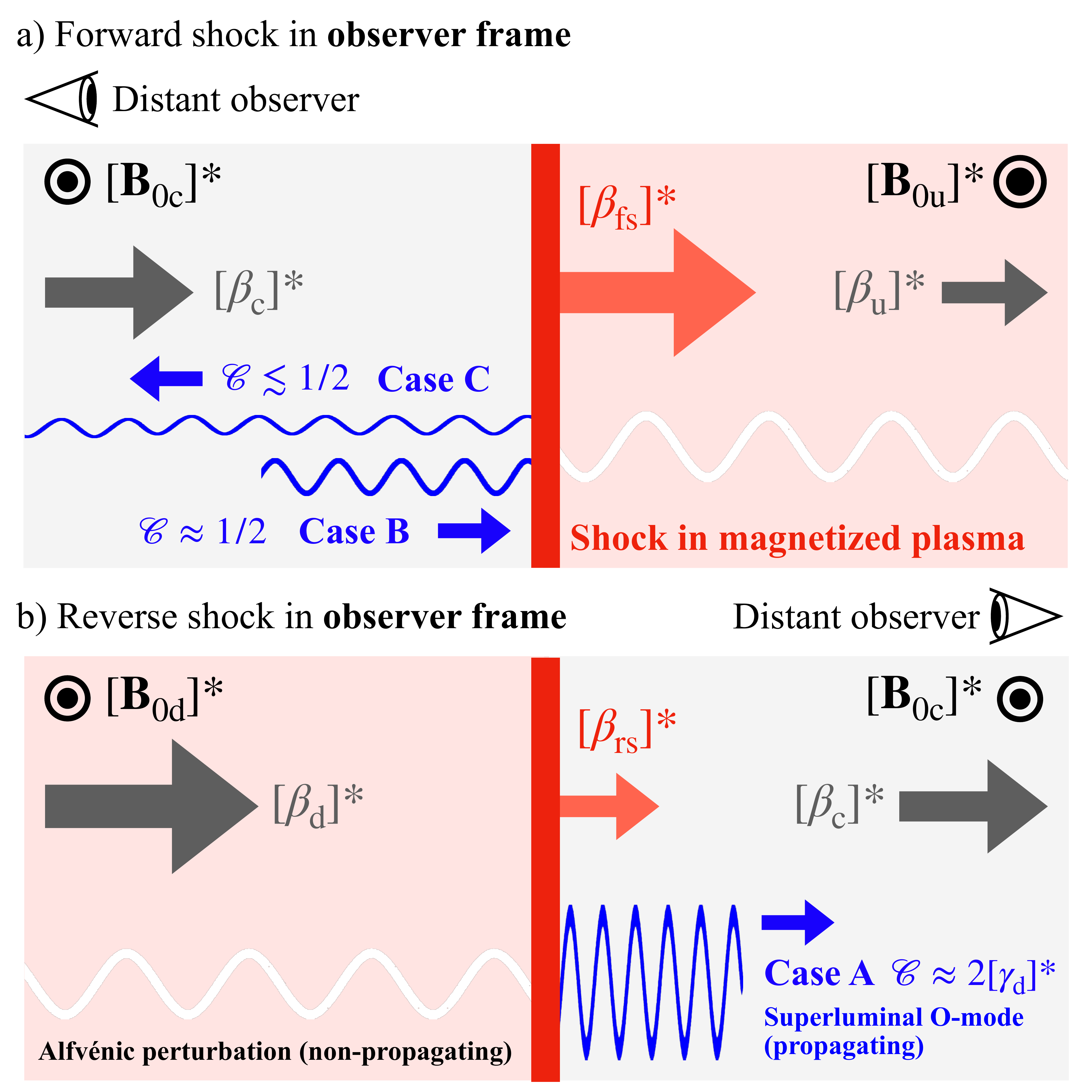}
    \vspace{-16pt}
    \caption{Schematic visualization of an observer frame's view of superluminal O-mode generation via Alfvén wave activation at magnetized shocks. A forward shock (panel a) and reverse shock (panel b) move through a propagating flow, like an expanding magnetar wind, as discussed in Appendix~\ref{app:ShockStructure}. Depending on the wave propagation direction in the frame of the contact discontinuity, amplitude and frequency experience different boosts $\mathcal{C}$, see also Section~\ref{sec:discussion}.}
    \label{fig:figure5}
    \end{figure}

For the following discussion, we analyze propagating waves with a frequency $\omega_{\rm O}$ and group velocity $\beta_{\rm O}$ in the downstream frame. A detector aligned with the shock normal observing the flow with a velocity $[\beta_{\rm d}]^\ast$ (see Figure~\ref{fig:figure1}) measures superluminal O-modes propagating with group velocity and frequency
\begin{align}
\begin{split}
    [\beta_{\rm O}]^\ast=\frac{[\beta_{\rm O}]^{\rm d}+[\beta_{\rm d}]^\ast}{1+[\beta_{\rm O}]^{\rm d}[\beta_{\rm d}]^\ast},
\end{split}\\
\begin{split}
    [\omega_{\rm O}]^\ast=[\gamma_{\rm d}]^\ast \left(1+[\beta_{\rm O}]^{\rm d}[\beta_{\rm d}]^\ast\right)[\omega_{\rm O}]^{\rm d}.
\end{split}
\end{align}
Their amplitude transforms to the observer frame as
\begin{align}
    [\delta E_{y,\rm d}]^\ast=[\gamma_{\rm d}]^\ast \left(1+[\beta_{\rm O}]^{\rm d}[\beta_{\rm d}]^\ast\right)[\delta E_{y,\rm d}]^{\rm d}.
\end{align}
For $[\beta_{\rm d}]^\ast$ along the $+\mathbf{\hat{x}}$ direction, we distinguish three possible cases of observed superluminal O-modes. 

\textit{Case A: }When $[\beta_{\rm O}]^{\rm d}\approx 1$ aligns with the direction of the downstream flow $[\beta_{\rm d}]^\ast$, then O-modes propagate along the flow direction with $[\beta_{\rm O}]^\ast\approx 1$ and experience a frequency and amplitude boost with a factor $\mathcal{C}=[\delta E_{y,\rm d}]^\ast/[\delta E_{y,\rm d}]^{\rm d}=[\omega_{\rm O}]^\ast/[\omega_{\rm O}]^{\rm d}\approx 2[\gamma_{\rm d}]^\ast$.

\textit{ Case B:} When $[\beta_{\rm O}]^{\rm d}$ points opposite to the flow with $[\beta_{\rm d}]^\ast$ and $[\beta_{\rm O}]^{\rm d}+[\beta_{\rm d}]^\ast>0$, then frequency and amplitude decrease by approximately $\mathcal{C}\approx 1/2$.

\textit{Case C:} When $[\beta_{\rm O}]^{\rm d}$ points opposite to the flow with $[\beta_{\rm d}]^\ast$ and $[\beta_{\rm O}]^{\rm d}+[\beta_{\rm d}]^\ast<0$, then frequency and amplitude decrease as $\mathcal{C}\approx (1/2)\times [\gamma_{\rm d}]^\ast/[\gamma_{\rm O}]^{\rm d}\lesssim 1/2$.

Appendix~\ref{app:ShockStructure} outlines a simplistic shock structure expected from collisions of homogeneous and highly magnetized ejecta shells \citep{Beloborodov2020,Thompson2022}. In such a scenario, the most favorable enhancement of frequencies and amplitudes corresponds to waves generated at the reverse shock (\textit{Case A}, see also Figure~\ref{fig:figure5}). For downstream waves with $[\omega_{\rm u}^2]^{\rm s}\gg\omega_{\rm pd}^2$, wave activation approximately enhances the frequency compared to the upstream perturbation by
\begin{align}
    [\omega_{\rm O}]^\ast\approx 2\times [\gamma_{\rm d}]^\ast [\gamma_{\rm u}]^{\rm d} [ck_{\rm u}]^{\rm u}\gtrsim 4\times[\gamma_{\rm u}]^\ast [ck_{\rm u}]^{\rm u}.
    \label{eq:reverseshockboost}
\end{align}
We evaluate astrophysical implications of such frequency boosts in the observer frame in the following section.

\subsection{Astrophysical Implications: Fast Radio Bursts}

The luminosity of superluminal O-modes generated by `shock-activation' depends on the available energy in upstream fluctuations above the critical scale given by Equation~(\ref{eq:downstreamcritical}). The fraction of turbulent energy available at $k d_0 \sim 1$ may be very small for realistic spectra and scale separations and is the main source of uncertainty when applying our results to FRB sources. \citet{Thompson2022} discusses the applicability of the shock-activation model in magnetar magnetospheres. Seed fluctuations generated by activity at the stellar surface likely have spatial scales of the order of a fraction of the neutron star radius $R_\ast$, $\lambda_{\rm seed}\approx 0.1R_\ast\approx 10^5$cm. The plasma skin depth $d_{0\ast}$ in the immediate magnetar vicinity can be estimated for electron densities $n_e=\mathcal{M}n_{\rm GJ}$, where $n_{\rm GJ}=\Omega B_\ast/(2\pi e c)$ is the Goldreich-Julian (GJ) density for a magnetosphere rotating with angular frequency $\Omega$ and surface magnetic field $B_\ast$. Magnetospheric activity like twists and crustal failures determine the multiplicity $\mathcal{M}$. We find typical values of  
\begin{align}
\begin{split}
    &d_{0\ast} = \left(\frac{m_e c^2}{4\pi n_e e^2}\right)^{1/2}\\
    &\approx 2\times10^{-3}\left(\frac{10^3}{\mathcal{M}}\right)^{1/2}\left(\frac{10^{15}{\rm G}}{B_\ast}\right)^{1/2}\left(\frac{P}{1{\rm s}}\right)^{1/2}\,{\rm cm}.
\end{split}
\end{align}
Thus, we expect initial seed perturbations with $d_{0\ast} k_\ast=2\pi d_{0\ast}/\lambda_{\ast}\lesssim 10^{-7}$. This is likely an upper limit as we expect charge densities above $\mathcal{M}n_{\rm GJ}$ during magnetospheric activities. Modeling the product $k\,d_{0}$ through the magnetosphere and wind zone is not straightforward. The skin depth $d_{0}$ depends on plasma properties like pair loading, bulk flow velocity, and temperature, varying substantially between different stages of magnetar activity and magnetospheric models. For example, if expanding in a quasi-thermal pair fireball, the rest-frame skin depth scales with radius, roughly $d_0\propto r/R_\ast$ \citep[cf.][]{Thompson2022}. Thus, to exceed the cutoff in Equation~(\ref{eq:downstreamcritical}), the seed waves must cascade over several decades in $k$, reaching $k\,d_{0}\!\approx\!1$.
For a cascading spectrum (Appendix~\ref{app:spectra}), the energy available at relevant FRB-inducing scales may be limited. For an energy scaling as $e(k)\propto k^{-s}$ for $k>k_\ast$, the fraction of available energy above the scale $k_{\rm seed}$ is
\begin{align}
    \frac{e(k>k_{\rm seed})}{e(k>k_\ast)}=\left(\frac{k_\ast}{k_{\rm seed}}\right)^{s-1}.
\end{align}
For instance, $k_{\rm seed}=10^3k_\ast$ and $s=5/3$ result in only $\lesssim 1\%$ of the wave energy available for generating propagating waves at the shock with $k>k_\ast$. The small fraction of turbulent energy at $k d_0 \sim 1$ makes quantitative predictions for FRBs uncertain.

Observed radio wave properties, like the escaping energy and overall frequency range, could inform magnetosphere models for compact objects like magnetars. Combining seed wave properties with efficiency estimates of transporting fluctuations to the shock location can inform plasma models of these environments. The low-frequency limit on emerging waves (Equation~\ref{eq:finalcutoff}) can constrain the plasma magnetization. For the reverse shock scenario (\textit{Case A}), Equations~(\ref{eq:downstreamcritical}) and~(\ref{eq:reverseshockboost}) estimate a cutoff of the observed frequency:
\begin{align}
    [\nu_O]^\ast \gtrsim\frac{2}{\pi}[\gamma_{\rm u}]^\ast \omega_{\rm pu}\left(\mathcal{S}[\sigma_{\rm u}]^{\rm u}\right)^{-1/2}.
    \label{eq:frequencygame}
\end{align}
FRBs typically cover narrowband frequency intervals of a few hundred MHz, in a space between 120MHz and several GHz \citep{PastorMarazuela2021}. This range allows for significant variations in the plasma properties contributing to the right-hand side of Equation~(\ref{eq:frequencygame}). The complex parameter space requires careful analysis and modeling. The potential effects of baryon loading, particularly in the context of fireball models \citep{bransgrove2025}, should also be evaluated.


Activation of Alfvénic perturbations to propagating superluminal modes can apply to other relativistic shocks in magnetically active environments. For magnetars, the plasma properties of relativistic outflows have a large parameter space \citep[e.g.,][]{Margalit2018,Beloborodov2020,Thompson2022}. Within the light cylinder, large-amplitude fast magnetosonic waves can shock when their electric field becomes comparable to the magnetic field \citep{chen2022,Beloborodov2023,Mahlmann2024,Bernardi2025,Vanthieghem2025}. Magnetar winds can also form shocks with companion stars in binary systems, possibly associated with periodically repeating radio transients \citep[e.g.,][]{Lyutikov2020,Barkov2022,Wang2022,Barkov2024,Wei2024}. The described wave activation mechanism has to be evaluated for typical plasma and shock strength parameters of these systems.

Theory for cold pair plasmas  (Section~\ref{sec:activation}) and 1D PIC simulations of linear, low-amplitude, seed perturbations (Section~\ref{sec:simulations}) demonstrate the validity of the wave generation mechanism discussed by \citet{Thompson2022}. Future studies have to evaluate several aspects of this mechanism, like efficiency of wave activation for varying upstream flow properties. The activation criterion discussed in this letter applies to Alfvénically polarized, nonpropagating magnetic shear fluctuations that are advected into the shock. Generalizing to 2D/3D will give crucial insights. If the upstream fluctuations were propagating AWs with finite $k_\parallel$, the associated electric fields could affect the mode matching across the shock and add resonant interactions with the shock structure. The activation condition derived here should therefore be viewed as the nonpropagating limit and a conservative threshold for more general upstream conditions.
Fluctuations at the shock could result in oblique superluminal modes ($k_\parallel\neq 0$) downstream of the shock. Turbulence fluctuations on scales of the downstream-propagating modes could play an important observational role \citep{Thompson2022}. Density blobs or magnetic structures could scatter activated superluminal modes and thereby change the properties measured in the observer frame. 
The coupling between shock-generated (superluminal) plasma waves and downstream turbulence requires numerical and theoretical investigation.

\citet{Plotnikov2019} and \citet{Sironi2021} analyzed kinetic plasma simulations of shock maser emission. This coherent process generates radiation, possibly at FRB frequencies. However, its efficiency decreases for increasing magnetization and depends on upstream conditions like temperature \citep{Babul2020,Vanthieghem2025a}. Future parameter scans of the wave activation mechanism discussed in this work will evaluate the simultaneous occurrence of synchrotron maser emission and mode activation. In the transition from mild ($\sigma\approx 1$) to strong magnetizations, wave activation at the shock may alleviate the efficiency constraints of precursor emission \citep{Thompson2022}.

This letter is a first-principles proof of concept for a mechanism that activates nonpropagating Alfvénic perturbations at relativistic magnetized shocks \citep{Thompson2022}. Small-scale fluctuations at scales above a critical wavenumber pass through the shock, generating propagating superluminal modes downstream. Depending on the flow and plasma properties, the activated waves could reach GHz frequencies as a possible source of FRBs. 


\begin{acknowledgements}

We are grateful for the anonymous referee's helpful comments. We thank Ludwig Böss, Hayk Hakobyan, Amir Levinson, Alexander Philippov, Siddhant Solanki, Anatoly Spitkovsky, and Christopher Thompson for valuable discussions. We thank Nicole Wotring for supporting the science communication efforts related to this work by producing animations based on the simulation data. L.E. and N.W. are grateful for the support by Dartmouth College's SURFD undergraduate research programs. J.F.M. acknowledges support from NSF grant AST-2508744. L.S. acknowledges support from the DOE Early Career Award DE-SC0023015, NASA ATP 80NSSC24K1826, NSF AST-2307202, and the Simons Foundation (MP-SCMPS-00001470). This work was facilitated by the Multimessenger Plasma Physics Center (NSF PHY-2206609) and supported in part by NSF PHY-2309135 to the Kavli Institute for Theoretical Physics. We acknowledge GitHub Copilot for coding assistance, and generative AI tools for figure design. Simulations were performed on the \textit{Ginsburg} (Columbia University), \textit{Rusty} (Flatiron Institute), and \textit{Discovery} (Dartmouth College) clusters. This work was granted access to the HPC resources of TGCC/CCRT under the allocation A0180415130 made by GENCI.
\end{acknowledgements}

%
%

\appendix

\section{Flow and wave properties in different frames}
\label{app:boosts}

\subsection{Relevant Frame Transformations and Identities}
\label{app:frames}

Table~\ref{tab:frames} shows a summary of frame transformations relevant for this manuscript. We commonly apply Lorentz boosts along the $x$-direction and their inverse for frequency and wavenumber measurements in the different frames. Primed quantities are measured in a frame moving at a relative velocity $\beta$.
\begin{align}
\omega' = \gamma\left(\omega-c\beta k\right)&\qquad k' =\gamma\left(k-\beta\omega/c\right)\,,\\
    \omega = \gamma(c\beta k^\prime + \omega^\prime)&\qquad k = \gamma\left(\beta\omega^\prime/c + k^\prime \right)\,.
\end{align}
Equation~(\ref{eq:omode}) provides the dispersion relation for superluminal O-modes in the plasma rest-frame. We derive the corresponding dispersion relation in the shock frame (s) by substituting primed quantities into Equation (\ref{eq:omode}):
\begin{equation}
    [\omega_{O\rm d}]^{\rm d} = [\gamma_{\rm d}]^{\rm s} \left(c[\beta_{\rm d} k_{\rm d}]^{\rm s} + [\omega_{O\rm d}]^{\rm s}\right) = \left[\omega_{\rm pd}^2 + c^2[\gamma_{\rm d}^2]^{\rm s} \left([k_{\rm d}]^{\rm s} + [\beta_{\rm d} \omega_{O\rm d}]^{\rm s}/c\right)^2\right]^{1/2}
\end{equation}
Rearranging this equation yields
\begin{equation}
    [\gamma_{\rm d}^2]^{\rm s} \left(c^2[k_{\rm d}^2(\beta^2_{\rm d} - 1)]^{\rm s} + [\omega_{O\rm d}^2(1-\beta_{\rm d}^2)]^{\rm s}\right) = \omega_{\rm pd}^2.
\end{equation}
Applying the identity $1-\beta^2 = 1/\gamma^2$, we obtain the O-mode dispersion in the shock front frame:
\begin{equation}
    [\omega_{O\rm d}]^{\rm s} = \left(c^2[k_{\rm d}^2]^{\rm s} + \omega_{\rm pd}^2\right)^{1/2}\label{eq:omodeshock}
\end{equation}

One instrumental identity is the ratio of upstream and downstream shock velocities $[\gamma_{\rm s}\beta_{\rm s}]^{\rm u}/[\gamma_{\rm s}\beta_{\rm s}]^{\rm d}$. We use the following velocity addition formulae:
\begin{align}
    [\gamma_{\rm s}]^{\rm u}&=[\gamma_{\rm s}]^{\rm d}[\gamma_{\rm u}]^{\rm d}\left(1-[\beta_{\rm s}]^{\rm d}[\beta_{\rm u}]^{\rm d}\right),\label{eq:shockboostgamma}\\
    [\beta_{\rm s}]^{\rm u}&=\frac{[\beta_{\rm s}]^{\rm d}-[\beta_{\rm u}]^{\rm d}}{1-[\beta_{\rm s}]^{\rm d}[\beta_{\rm u}]^{\rm d}}.\label{eq:shockboost}
\end{align}
Combining the above and considering large magnetization and relativistic upstream velocities, we find:
\begin{align}
    \frac{[\gamma_{\rm s}\beta_{\rm s}]^{\rm u}}{[\gamma_{\rm s}\beta_{\rm s}]^{\rm d}}=\left(1-\frac{[\beta_{\rm u}]^{\rm d}}{[\beta_{\rm s}]^{\rm d}}\right)[\gamma_{\rm u}]^{\rm d}\approx 2[\gamma_{\rm u}]^{\rm d}.\label{eq:shockvelfactor}
\end{align}
Another approximation of Equation~(\ref{eq:shockboostgamma}) in the limit of high flow velocities in the shock frame, $[\beta_{\rm d}]^{\rm s}\approx -1$ and $[\beta_{\rm s}]^{\rm u}\approx 1$, is
\begin{align}
    [\gamma_{\rm d}]^{\rm s}[\gamma_{\rm s}]^{\rm u}=\frac{[\gamma_{\rm d}]^{\rm u}}{1-[\beta_{\rm d}]^{\rm s}[\beta_{\rm s}]^{\rm u}}\approx\frac{[\gamma_{\rm d}]^{\rm u}}{2}\label{eq:Oshockvelfactor}.
\end{align}
For superluminal O-modes, Equation~(\ref{eq:omodematching}) determines the downstream wavenumber in the shock frame. These scales can be further boosted into the downstream frame, where
\begin{align}
    [k_{\rm d}]^{\rm d} = [\gamma_{\rm d}]^{\rm s}\left([k_{\rm d}]^{\rm s} - [\beta_{\rm d}]^{\rm s}[\omega_{\rm d}]^{\rm s}/c\right)=[\gamma_{\rm d}]^{\rm s}\left(\frac{1}{c}(c^2[\gamma_{\rm s}^2\beta_{\rm s}^2 k_{\rm u}^2]^{\rm u} - \omega_{\rm pd}^2)^{1/2} - [\beta_{\rm d}]^{\rm s}[\gamma_{\rm s}\beta_{\rm s} k_{\rm u}]^{\rm u}\right)\,.
\end{align}
In the limit of $[\omega_{\rm d}]^{\rm s} \gg \omega_{\rm pd}$ and $[\beta_{\rm d}]^{\rm s}\approx -1$, this expression simplifies:
\begin{align}
    [k_{\rm d}]^{\rm d} \approx [\gamma_{\rm d}]^{\rm s}[\gamma_{\rm s} \beta_{\rm s} k_{\rm u}]^{\rm u}\left(1 - [\beta_{\rm d}]^{\rm s}\right) \approx 2[\gamma_{\rm d}]^{\rm s}[\gamma_{\rm s} \beta_{\rm s} k_{\rm u}]^{\rm u}\approx [\gamma_{\rm u}]^{\rm d}[\beta_{\rm s} k_{\rm u}]^{\rm u}\,.
    \label{eq:kddscale}
\end{align}
In the last step, we used the velocity addition formula assuming relativistic shock velocities: $2[\gamma_d]^s[\gamma_s]^u \approx [\gamma_d]^u = [\gamma_u]^d$. We can express Equation~(\ref{eq:kddscale}) as $[k_{\rm u}]^{\rm d} = [k_{\rm d}]^{\rm d}/[\beta_{\rm s}]^{\rm u}$.

\begin{table}[t]
\centering
\caption{Key quantities measured in different reference frames.}
\label{tab:frames}
\renewcommand{\arraystretch}{1.15}

\begin{tabularx}{\textwidth}{
    p{0.175\textwidth}  
    >{\arraybackslash}m{0.175\textwidth}
    >{\arraybackslash}m{0.175\textwidth}
    >{\arraybackslash}m{0.175\textwidth}
    >{\arraybackslash}m{0.3\textwidth}
}
\toprule
\textbf{Quantity} & \textbf{Downstream (d)} & \textbf{Shock front (s)} & \textbf{Upstream (u)} & \textbf{Notes} \\
\midrule
$\omega_{A\rm u}$ & $c\left[\gamma_{\rm u}\beta_{\rm u}\right]^{\rm d}\,\left[k_{\rm u}\right]^{\rm u}$  &  $-c\left[\gamma_{\rm s}\beta_{\rm s}k_{\rm u}\right]^{\rm u}$ & ${0}$ & Applied in Equation~(\ref{eq:udjump}) \\
$\omega_{A\rm d}$ & ${0}$ & $-c\left[\gamma_{\rm s}\beta_{\rm s} k_{\rm d}\right]^{\rm d}$ & $c\left[\gamma_{\rm u}\beta_{\rm u}\right]^{\rm d}\,\left[k_{\rm d}\right]^{\rm d}$ & Applied in Equation~(\ref{eq:udjump}) \\
$\omega_{O\rm d}$ & $\left(c^2[k_{\rm d}^2]^{\rm d} + \omega_{\rm pd}^2\right)^{1/2}$ & $\left(c^2[k_{\rm d}^2]^{\rm s} + \omega_{\rm pd}^2\right)^{1/2}$ & $\left(c^2[k_{\rm d}^2]^{\rm u} + \omega_{\rm pd}^2\right)^{1/2}$ & Derived in Equation~(\ref{eq:omodeshock}) \\
\midrule
$\delta B_{z\rm u}$ & $[\gamma_{\rm d} \delta B_{\rm u}]^{\rm u}$ & $[\gamma_{\rm s} \delta B_{\rm u}]^{\rm u}$ & $[\delta B_{u}]^{\rm u}$ & For $[\delta \bm{E}_{\rm u}]^{\rm u} = \boldsymbol{0}$ \\
$\delta E_{y\rm u}$ & $-[\gamma_{\rm d} \beta_{\rm d} \delta B_{\rm u}]^{\rm u}$ & $-[\gamma_{\rm s}\beta_{\rm s} \delta B_{\rm u}]^{\rm u}$ & $0$ & \\
$\delta B_{z\rm d}$ & $[\delta B_{\rm d}]^{\rm d}$ & $[\gamma_{\rm s}\delta B_{\rm d}]^{\rm d}$ & $[\gamma_{\rm u}\delta B_{\rm d}]^{\rm d}$ & For $[\delta \bm{E}_{\rm d}]^{\rm d} = \boldsymbol{0}$ \\
$\delta E_{\rm yd}$ & $0$ & $-[\gamma_{\rm s} \beta_{\rm s} \delta B_{\rm d}]^{\rm d}$ & $-[\gamma_{\rm u} \beta_{\rm u} \delta B_{\rm d}]^{\rm u}$ &  \\
\bottomrule
\bottomrule
\end{tabularx}
\end{table}

\subsection{Shock structure from homogeneous and strongly magnetized shell collisions}
\label{app:ShockStructure}

The astrophysical interpretation of the mode conversion mechanism evaluated in this letter relies on the shock structure induced during collisions of uniform and highly magnetized shells discussed by \citet[][Section 6]{Thompson2022} and, more generally, by \citet{Beloborodov2020}. We consider a structure of reverse shock (rs), contact discontinuity (c), and forward shock (fs), propagating through a relativistic wind with an upstream (u) velocity $\gamma_{u|\ast}$ and a downstream (d) velocity $\gamma_{d|\ast}$. The specific solution discussed here has $\gamma_{{\rm rs}|\ast}\ll\gamma_{{\rm c}|\ast}\ll\gamma_{{\rm fs}|\ast}$. The MHD jump conditions imply continuous tangential electric fields in the shock-front-frame:
\begin{align}
    E_{{\rm c|fs}}^z=E_{{\rm u|fs}}^z\label{eq:JumpEfs}
\end{align}
We boost the corresponding lab-frame fields by applying the following transformations and using $\mathbf{E}=-\boldsymbol{\beta}\times\mathbf{B}$:
\begin{align}
    E^z_{\rm u|fs}=\gamma_{\rm fs|\ast}\left(E^z_{\rm u|\ast}+\beta_{\rm fs|\ast} B^y_{\rm u|\ast}\right)=\gamma_{\rm fs|\ast}\left(-\beta_{\rm u|\ast}+\beta_{\rm fs|\ast}\right)B^y_{\rm u|\ast}\approx\gamma_{\rm fs|\ast}\left(1-\beta_{\rm u|\ast}\right)B^y_{\rm u|\ast}.
\end{align}
Here, we used that $\beta_{\rm u|\ast}<\beta_{\rm fs|\ast}\approx 1$. Equation~(\ref{eq:JumpEfs}) can be simplified as follows: 
\begin{align}
    B_{{\rm c|\ast}}^y=\left(\frac{1-\beta_{\rm u|\ast}}{1-\beta_{\rm c|\ast}}\right) B_{{\rm u|\ast}}^y=\left(\frac{1-\beta_{\rm u|\ast}^2}{1-\beta_{\rm c|\ast}^2}\right)\left(\frac{1+\beta_{\rm c|\ast}}{1+\beta_{\rm u|\ast}}\right) B_{{\rm u|\ast}}^y\approx \frac{\gamma_{\rm c|\ast}^2}{\gamma_{\rm u|\ast}^2}B_{{\rm u|\ast}}^y.
\end{align}
In direct analogy, using $\beta_{\rm rs|\ast}<\beta_{\rm d|\ast}\approx 1$, the jump across the reverse shock is $B_{{\rm d|\ast}}^y\approx B_{{\rm c|\ast}}^y$. Combining these relations across the entire shock structure lets us write
\begin{align}
    \frac{\gamma_{\rm c|\ast}}{\gamma_{\rm u|\ast}}=\left(\frac{B_{{\rm d|\ast}}^y}{B_{{\rm u|\ast}}^y}\right)^{1/2}=\left(\frac{L_{{\rm d|\ast}}}{L_{{\rm u|\ast}}}\right)^{1/4},
\end{align}
where $L$ are the spherical Poynting luminosities associated with a shell. It is instructive to define the shock strength parameters as the ratio of the flow Lorentz factors in the respective shock's frame:
\begin{align}
    \text{(FS)}\qquad \mathcal{S}_{\rm fs}&=\frac{\gamma_{\rm u|fs}}{\gamma_{\rm c|fs}}=\frac{\gamma_{\rm fs|\ast}/2\gamma_{\rm u|\ast}}{\gamma_{\rm fs|\ast}/2\gamma_{\rm c|\ast}}=\frac{\gamma_{\rm c|\ast}}{\gamma_{\rm u|\ast}}=\left(\frac{L_{{\rm d|\ast}}}{L_{{\rm u|\ast}}}\right)^{1/4}>1\\
    \text{(RS)}\qquad \mathcal{S}_{\rm rs}&=\frac{\gamma_{\rm d|rs}}{\gamma_{\rm c|rs}}=\frac{\gamma_{\rm d|\ast}/2\gamma_{\rm rs|\ast}}{\gamma_{\rm c|\ast}/2\gamma_{\rm rs|\ast}}=\frac{\gamma_{\rm d|\ast}}{\gamma_{\rm c|\ast}}=\frac{\gamma_{\rm d|\ast}}{\mathcal{S}_{\rm fs}\gamma_{\rm u|\ast}}=\frac{\gamma_{\rm d|\ast}}{\gamma_{\rm u|\ast}}\left(\frac{L_{{\rm u|\ast}}}{L_{{\rm d|\ast}}}\right)^{1/4}>1.
\end{align}
We use the MHD jump conditions further to obtain flow properties in the respective shock frames:
\begin{align}
    \text{(FS)}\qquad &\gamma_{\rm u|fs}=\left[\sigma_{\rm u}\left(\frac{1}{2}+\mathcal{S}_{\rm fs}^2\right)\right]^{1/2}\qquad \gamma_{\rm c|fs}=\frac{\gamma_{\rm u|fs}}{\mathcal{S}_{\rm fs}}\label{eq:gammaufs},\\
    \text{(RS)}\qquad &\gamma_{\rm d|rs}=\left[\sigma_{\rm d}\left(\frac{1}{2}+\mathcal{S}_{\rm rs}^2\right)\right]^{1/2}\qquad \gamma_{\rm c|rs}=\frac{\gamma_{\rm d|fs}}{\mathcal{S}_{\rm rs}}\label{eq:gammadrs}.
\end{align}
Finally, we can obtain the upstream velocities in the frame of the contact discontinuity for both shocks:
\begin{align}
    \text{(FS)}\qquad & \beta_{\rm u|c}=\frac{1-\mathcal{S}_{\rm fs}^2}{1+\mathcal{S}_{\rm fs}^2}\qquad \gamma_{\rm u|c} = \frac{1}{2}\left(\frac{1}{\mathcal{S}_{\rm fs}}+\mathcal{S}_{\rm fs}\right)\label{eq:gammauc}\\
    \text{(RS)}\qquad & \beta_{\rm d|c}=\frac{\mathcal{S}_{\rm rs}^2-1}{1+\mathcal{S}_{\rm rs}^2}\qquad \gamma_{\rm d|c} = \frac{1}{2}\left(\frac{1}{\mathcal{S}_{\rm rs}}+\mathcal{S}_{\rm rs}\right)\label{eq:gammadc}
\end{align}

\subsection{Jump Conditions}
\label{app:jumps}

 Continuity, momentum, and energy conservation in ideal relativistic MHD provide shock jump conditions. In the shock front frame, we find \citep[e.g.,][]{Lemoine2016}:
    \begin{align}
        [[J^\alpha n_\alpha]] = 0 \qquad
        [[T^{\alpha\beta}n_\alpha]] = 0 \qquad
        [[^*F^{\alpha\beta}n_\alpha]] = 0\,.
    \end{align}
    Here, double brackets $[[\dots]]$ denote the difference of a quantity between  immediate downstream and upstream. The normal vector in the shock front frame is $n_\alpha = [0, 1, 0, 0]$. $J^\alpha = \rho u^\alpha$ is the four-current in terms of the flow density $\rho$ and shock four-velocity $u^\alpha = [\gamma_{\rm s}, \gamma_{\rm s}\beta_{\rm s}, 0, 0]$, $T^{\alpha\beta}$ refers to the energy-momentum tensor. $^*F^{\alpha\beta}$ is the (dual) electromagnetic strength tensor, commonly expressed in terms of the shock velocity and magnetic field four-vector $^*F^{\alpha\beta} = u^\alpha b^\beta - u^\beta b^\alpha$.
    For perpendicular shocks (Section~\ref{sec:setup}) we find the following four conserved quantities across the shock:
    \begin{equation}
        [[\beta_{\rm s} B_0]]^{\rm s} = 0
    \end{equation}
    \begin{equation}
        [[\rho\gamma_{\rm s}\beta_{\rm s}]]^{\rm s} = 0
    \end{equation}
    \begin{equation}
        \left[\left[\left(w+\frac{B_0^2}{4\pi\gamma_{\rm s}    ^2}\right)\gamma_{\rm s}^2\beta_{\rm s}  \right]\right]^{\rm s} = 0
    \end{equation}
    \begin{equation}
        \left[\left[\left(w+\frac{B_0^2}{4\pi\gamma_{\rm s}    ^2}\right)\gamma_{\rm s}^2\beta_{\rm s}^2 + \left(p+\frac{B_0^2}{8\pi\gamma_{\rm s}^2}\right)  \right]\right]^{\rm s} = 0
    \end{equation}
    For strong shocks, like those considered throughout this letter, we can approximate $\gamma_{\rm s} \approx \sqrt{\sigma}$. 
    

\section{Mode activation for spectra of seed waves}
\label{app:spectra}

\begin{figure*}
\includegraphics[width=0.95\textwidth]{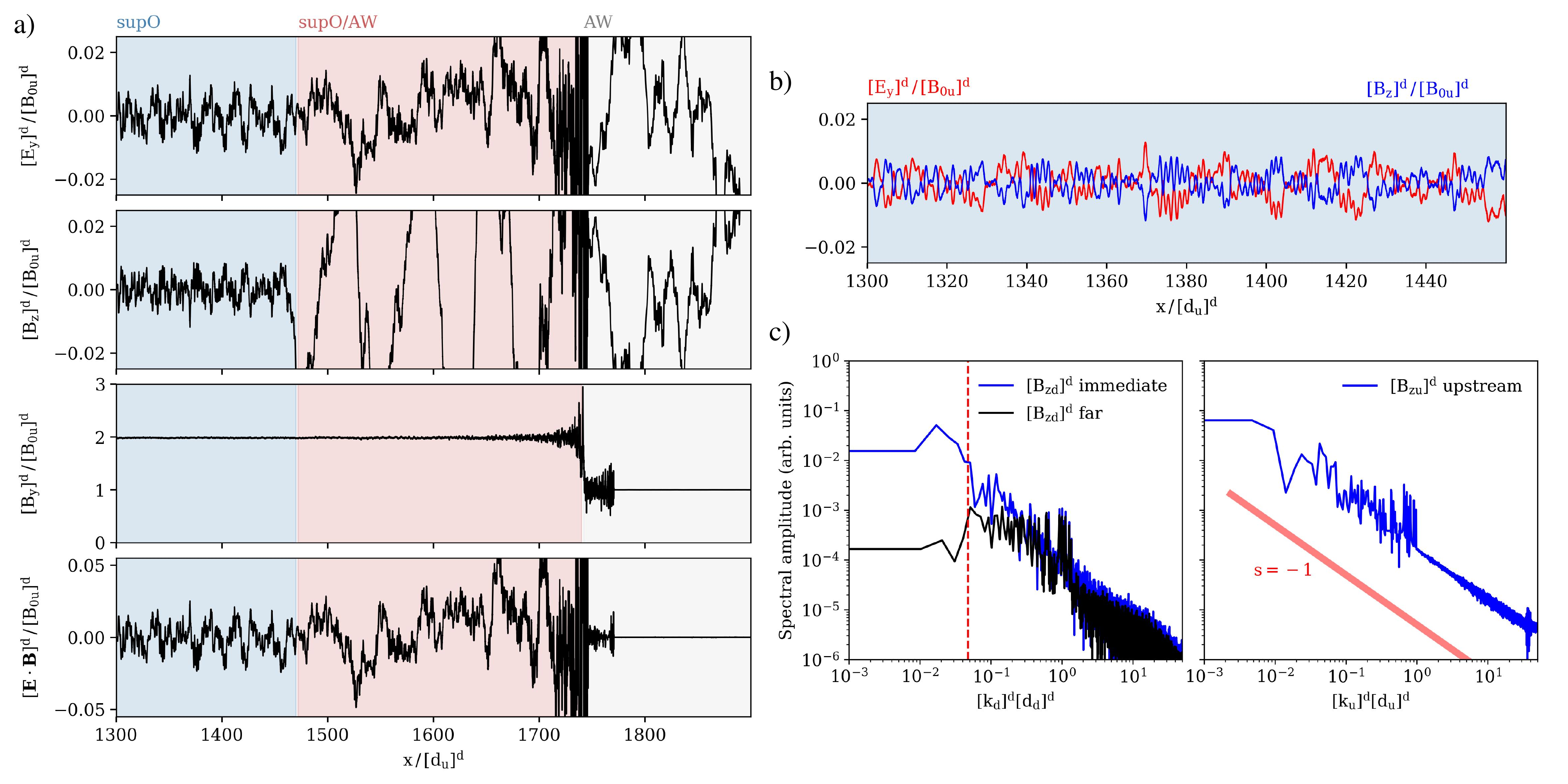}
\vspace{-6pt}
\caption{Same as Figure~\ref{fig:MSHOCK_G4_s25_HR_sPEC10}, but for a spectrum of seed perturbations with $s=1$ sampled with $N=100$ modes. We show the field components (panel a), a zoom of the wave EM fields (panel b), and a Fourier spectrum of the wave magnetic field in different regions (panel c). The Fourier spectrum distinguishes the immediate and far downstream (c, left panel), and the seed perturbations upstream of the shock (c, right panel). The red dashed line estimates the cutoff (cold plasma estimate, Equation~\ref{eq:mink}) for downstream scales, as outlined in Appendix~\ref{app:spectra}. The plasma frequency cutoff acts like a high-pass filter for waves in the far downstream.}
\label{fig:FIGURE_A1}
\end{figure*}

We extend the analysis of Section~\ref{sec:spectrumAW} to a spectrum of seed waves sampled from an energy distribution with decay index $s=1$. Figure~\ref{fig:FIGURE_A1} shows the corresponding wave activation dynamics. As outlined in Figure~\ref{fig:MSHOCK_G4_s25_HR_sPEC10}, seed waves at scales above the cutoff (Equation~\ref{eq:mink}) are activated and propagate far downstream of the shock. From Equation~(Equation~\ref{eq:mink}) and a suitable length contraction of the wave scale, we estimate the downstream cutoff in the cold plasma limit as
\begin{align}
    \frac{c[k_{\rm d}]^{\rm d}}{\omega_{\rm pd}}\gtrsim \frac{\left([\gamma_{\rm u}]^{\rm d}\right)^{1/2}}{[\gamma_{\rm s}]^{\rm u}}.
    \label{eq:downlimit}
\end{align}
Figure~\ref{fig:FIGURE_A1} (panel c) evaluates Fourier spectra of upstream and downstream wave scales. Propagating modes roughly appear above the limit estimated in Equation~(\ref{eq:downlimit}), indicated by a red dashed line. The plasma frequency cutoff acts like a high-pass filter with the overall energy distribution following the slope of the seed wave spectrum.

\bibliographystyle{aasjournal}
\bibliography{literature.bib}

\end{document}